%% file: main.tex
\documentclass[sigconf]{acmart}
\usepackage[utf8]{inputenc}
\usepackage[T1]{fontenc}

\usepackage{newtxmath} 
\usepackage{array}
\usepackage{multirow}
\usepackage{graphicx}
\usepackage{subfig}
\usepackage{svg}
\usepackage{amsmath}
\usepackage{balance}
\usepackage{multicol}
\usepackage[ruled,vlined,lined,commentsnumbered]{algorithm2e}
\usepackage{esvect}
\usepackage{booktabs}
\usepackage{tabularx}
\usepackage{anyfontsize}
\usepackage[all]{nowidow}
\usepackage[normalem]{ulem}
\usepackage{xcolor}
\usepackage{colortbl}
\usepackage{makecell}
\usepackage[switch]{lineno}
\usepackage{framed}
\usepackage{wrapfig}
\usepackage{varwidth}
\usepackage{lineno}
\usepackage{fancyvrb}
\usepackage{adjustbox}
\usepackage{pifont}
\usepackage{amsmath}

\usepackage{listings}

\usepackage{framed}
\usepackage{tikz}
\definecolor{light-gray}{gray}{0.9}
\usepackage[framemethod=TikZ]{mdframed}
\mdfsetup{skipabove=5pt,skipbelow=3pt}
\usepackage{lipsum}
\mdfdefinestyle{RQFrame}{%
	linecolor=black,
	outerlinewidth=0.15pt,
	roundcorner=3pt,
	innertopmargin=2pt,
	innerbottommargin=2pt,
	innerrightmargin=4pt,
	innerleftmargin=4pt,
	backgroundcolor=light-gray}

\newcommand{\app}[0]{MELA}

\input{Sections/headerpackages}

\newcommand{\system}[0]{RRTRouter}

\definecolor{commentGray}{RGB}{120,120,120}
\renewcommand{\algorithmiccomment}[1]{\bgroup\color{commentGray}{//#1}\egroup}

\definecolor{myBlue}{RGB}{0,100,230}
\newcommand\requirement[1]{%
    \tikz[baseline=(req.base)]\node [draw=myBlue, fill=myBlue, thick, rectangle, inner sep=2pt, rounded corners=2pt] (req) {\color{white}\textbf{#1}};
}
\title{Enhancing Automata Learning with Statistical Machine Learning: A Network Security Case Study}

\copyrightyear{2024}
\acmYear{2024}
\setcopyright{acmlicensed}\acmConference[MODELS '24]{ACM/IEEE 27th International Conference on Model Driven Engineering Languages and Systems}{September 22--27, 2024}{Linz, Austria}
\acmBooktitle{ACM/IEEE 27th International Conference on Model Driven Engineering Languages and Systems (MODELS '24), September 22--27, 2024, Linz, Austria}
\acmDOI{10.1145/3640310.3674087}
\acmISBN{979-8-4007-0504-5/24/09}

\begin{document}

\begin{abstract}
Intrusion detection systems are crucial for network security. Verification of these systems is complicated by various factors, including the heterogeneity of network platforms and the continuously changing landscape of cyber threats. In this paper, we use automata learning to derive state machines from network-traffic data with the objective of supporting behavioural verification of intrusion detection systems. The most innovative aspect of our work is addressing the inability to directly apply existing automata learning techniques to network-traffic data due to the numeric nature of such data. Specifically, we use interpretable machine learning (ML) to partition numeric ranges into intervals that strongly correlate with a system's decisions regarding intrusion detection. These intervals are subsequently used to abstract numeric ranges before automata learning. We apply our ML-enhanced automata learning approach to a commercial network intrusion detection system developed by our industry partner, RabbitRun Technologies.  Our approach results in an average 67.5\% reduction in the number of states and transitions of the learned state machines, while achieving an average 28\% improvement in accuracy compared to using expertise-based numeric data abstraction. Furthermore, the resulting state machines help practitioners in verifying system-level security requirements and exploring previously unknown system behaviours through model checking and temporal query checking. 
We make our implementation and experimental data available online.
\end{abstract}


\author{Negin Ayoughi}
\affiliation{%
  \institution{University of Ottawa, Canada}
}
\email{negin.ayoughi@uottawa.ca}

\author{Shiva Nejati}
\affiliation{%
  \institution{University of Ottawa, Canada}
}
\email{snejati@uottawa.ca}

\author{Mehrdad Sabetzadeh}
\affiliation{%
  \institution{University of Ottawa, Canada}
}
\email{m.sabetzadeh@uottawa.ca}

\author{Patricio Saavedra}
\affiliation{%
  \institution{RabbitRun Technologies Inc., Canada}
}
\email{pat@rabbit.run}

\maketitle

\input{Sections/intro}
\input{Sections/industry-cleaned}

\input{Sections/approach}

\input{Sections/eval}

\input{Sections/relwork}

\input{Sections/lessons}

\onecolumn \begin{multicols}{2}

\bibliographystyle{ACM-Reference-Format}
\bibliography{bibliography}
\end{multicols}
\end{document}

%% file: Sections/headerpackages.tex
\usepackage[inline]{enumitem}
\usepackage{booktabs}
\usepackage[most]{tcolorbox}
\usepackage{fancybox}
\usepackage{algpseudocode}
\usepackage{makecell}
\usepackage[inline]{enumitem}

\definecolor{myBlue}{RGB}{0,133,255}
\definecolor{myOrange}{RGB}{255,115,0}
\definecolor{myGreen}{RGB}{0,115,0}

\newboolean{showcomments}
\setboolean{showcomments}{true}
\ifthenelse{\boolean{showcomments}}
{\newcommand{\nb}[2]{
  \fcolorbox{black}{yellow}{\bfseries\sffamily\scriptsize#1}
  {\sf\small$\blacktriangleright$\textit{#2}$\blacktriangleleft$}
 }
 
}
{\newcommand{\nb}[2]{}
 
}

\definecolor{lightgray}{gray}{0.75}

\newcommand\real{\ensuremath{\mathbb{R}}}

\newcommand\inputs{\ensuremath{\overline{\texttt{i}}}}
\newcommand\inputsignal{\ensuremath{i}}

\newcommand\outputs{\ensuremath{\overline{\texttt{o}}}}
\newcommand\outputsignal{\ensuremath{o}}

\usepackage[switch]{lineno}

%% file: Sections/intro.tex
\section{Introduction}
\label{sec:intro}
Our work stems from the needs of our industry partner, RabbitRun Technologies (RRT for short). RRT develops affordable network routers for small office and home office (SOHO) users -- a customer base that has grown in importance and size during and post-pandemic. Network routers are complex devices that handle various protocols and services. RRT’s router is equipped with a network intrusion detection system that monitors network traffic for suspicious activities indicating potential attacks. An effective network intrusion detection system is particularly important for SOHO environments, which often lack extensive security infrastructure and dedicated IT personnel to consistently manage threats.

Understanding and characterizing system-level behaviours of a complex network intrusion detection system poses a challenge due to the heterogeneity of network platforms and the constantly evolving landscape of cyber threats. Developing such understanding and being able to precisely capture system-level behaviours are nonetheless paramount for our industry partner for several reasons, including (a) providing better support resources and guidelines for network administrators and operators, (b) improving the identification of security vulnerabilities, such as unexpected responses to certain types of network traffic, and (c) having an analyzable specification against which to test new router software versions. To facilitate the above, \emph{we investigate the feasibility and effectiveness of automatically learning system-level behavioural models, more specifically system-level automata, to capture the behaviours of RRT's network intrusion detection system.}

Automata learning is the process of inferring automata models from observed behaviour~\cite{Vaandrager17,muskardin2022active,muvskardin2022aalpy,GarhewalD23,NeiderSVK97}. When behavioural models are unavailable or incomplete, automata learning becomes an important tool for improving system understandability and enabling model-based analysis~\cite{GarhewalD23}.
 Automata learning can be done in either an active or passive mode. Active learning involves algorithms interacting with the system under learning to generate data, whereas passive learning uses existing datasets like log files. Active automata learning has been used in real-world scenarios, including network protocols like BLE~\cite{PferscherA22}, MQTT~\cite{tappler2017model}, and TCP~\cite{fiteruau2016combining}, where a clearly defined interface with the system under learning is available. Network intrusion detection systems, however, analyze the continuous and numeric properties of network-traffic flows over time to detect  attacks, such as denial of service~\cite{zargar2013survey}. That is, their inputs and outputs consist of time-series data involving numeric values. These systems cannot support an iterative query-and-response closed-loop, which is crucial for active learning. Passive learning, on the other hand, is applicable to network intrusion detection systems. However, passive learning alone cannot handle raw numeric time-series data due to the excessive number of distinct values. Abstracting numeric values into meaningful intervals or categories is therefore paramount for effectively applying passive learning in the context of network intrusion detection systems.
 
In this paper, we propose the \emph{MachinE Learning-enhanced passive Automata learning approach (\app)} to derive state machines for network intrusion detection systems with numeric time-series inputs and outputs. Our approach generates network flows simulating both normal and attack scenarios, and captures the system outputs.  We transform the time-series data obtained from the network flows and system outputs into a set of traces. This transformation involves partitioning raw, numeric ranges into a set of intervals. We use decision trees to optimize the correlation of the resulting intervals with a system's decisions regarding intrusion detection.  We then use  passive automata learning~\cite{cano2010inferring} to derive state machines that capture the behaviours of RRT's \hbox{network intrusion detection system.}

\textbf{Contributions.} Our paper demonstrates the effective application of passive automata learning for building accurate and practically useful state machines for systems with numeric time-series inputs and outputs. The key idea behind our approach is using decision-tree learners to abstract the numeric data in traces so as to improve the accuracy and precision of the learned state machines. We evaluate our approach, \app, using an industrial network intrusion detection system developed by our partner, RRT. 
We compare the state machines learned using \app\ with those derived using a baseline approach that employs automata learning alongside expertise-based data abstraction.

Our results indicate that, when given the same set of time-series data, \app\ generates state machines with an average of 67.5\% fewer transitions and states, while, on average, achieving 28\% higher accuracy compared to the baseline. That is, combining automata learning with our ML-driven data abstraction leads to more concise state machines that can more accurately represent the behaviours of the system under learning. Further, we show how, using temporal model checking~\cite{mcbook} and query checking~\cite{Chan00},  the learned state machines help our industry partner with verifying system-level requirements and identifying previously unknown system-level behaviours. 

%% file: Sections/industry-cleaned.tex
\section{Intrusion Detection Case Study}
\label{sec:testbed}
\begin{figure}[t]
    \centering
    \includegraphics[width=.85\linewidth]{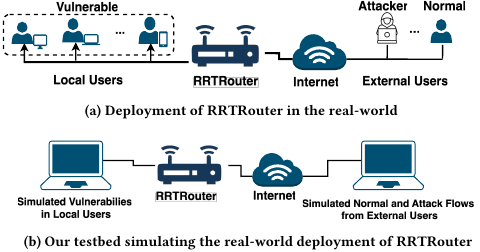}
      \vspace*{-.2cm}
   \caption{Our case study system: (a) Deployment of our industry partner's router (\system) in the real-world, and (b) our testbed simulating the real-world deployment.}
       \Description{This figure shows two images: (a) the deployment of our industry partner's router in the real world, and (b) a testbed simulating this real-world deployment.}
   \label{fig:fig_1}
       \vspace*{-.3cm}
\end{figure}

Figure~\ref{fig:fig_1}a illustrates the deployment of an RRT network router, which enables local users to connect to the Internet and external networks.  The router is equipped with an intrusion detection system to identify denial of service (DoS) and distributed denial of service (DDoS) attacks originating from external networks.  DoS attacks flood their target with excessive traffic from a single source, whereas DDoS attacks employ multiple sources to perform a more extensive attack~\cite{zargar2013survey}. 
RRT's intrusion detection system monitors the incoming traffic to the router for identifying attacks. If an attack is detected, the system moves to an alert state, triggering actions to block unauthorized traffic.  We refer to an RRT router (together with the intrusion detection system running on it) as \emph{\system}.

To test \system\  with different traffic flows, in collaboration with RRT, we have designed and implemented a testbed shown in Figure~\ref{fig:fig_1}b. The testbed enables the simulation of a varying number of external users, located to the right of \system, sending network flows to the local users, located to the left of \system.  As we discuss in Section~\ref{subsec: implementation}, the testbed uses a tool to simultaneously generate normal network flows alongside DoS and DDoS attacks from external users. It further uses a penetration testing framework to simulate vulnerabilities in local users. The network flows from external users pass through  \system, allowing it to collect data, analyze this data, and identify DoS / DDoS attacks.

\begin{figure}
	\centering
        \includegraphics[width=.85\linewidth]{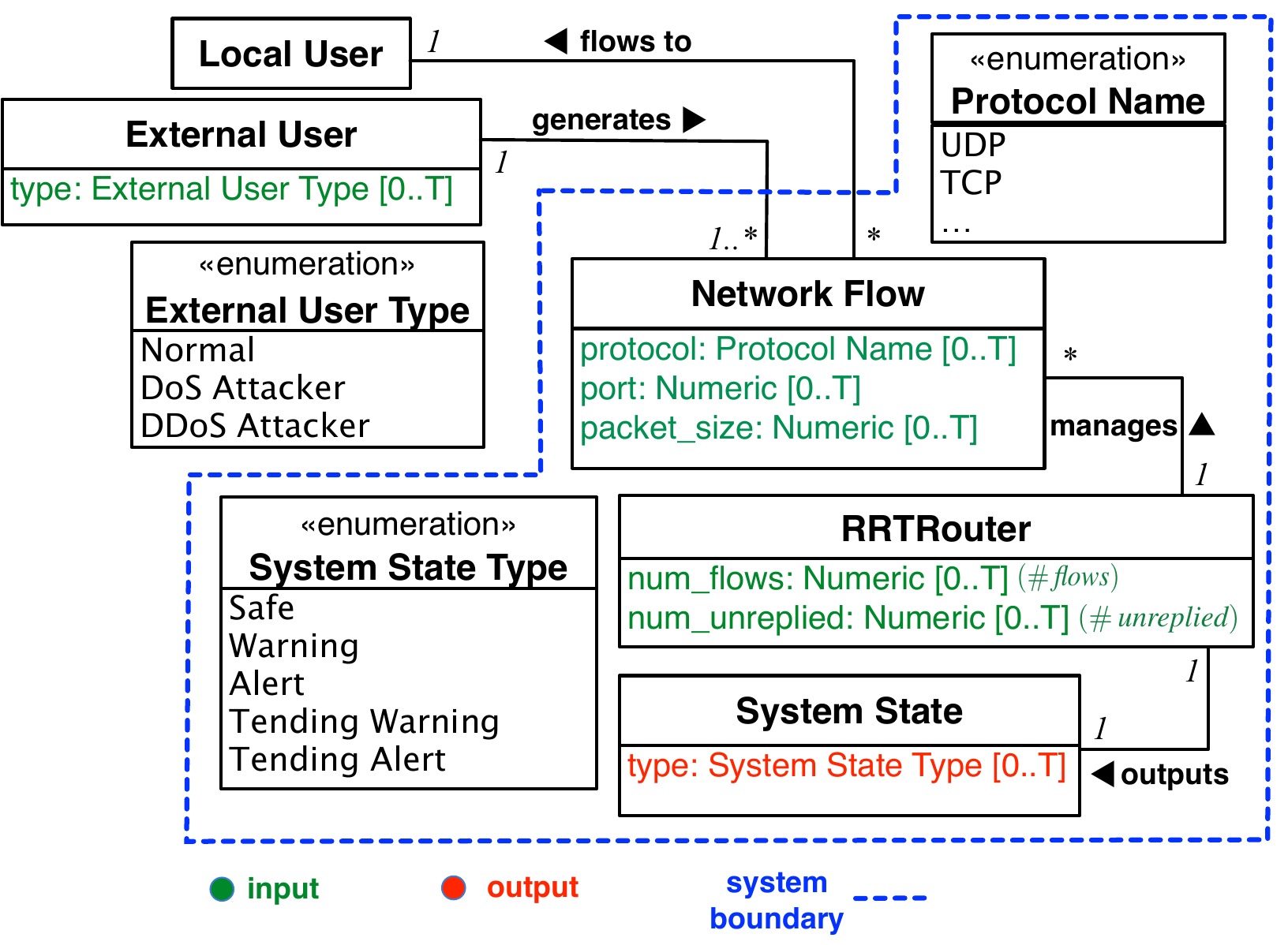}
        \vspace*{-.2cm}
        \caption{A conceptual model for our testbed.}
        \Description{A conceptual model for our testbed.}
		\label{fig:conceptual}
     \vspace*{-.3cm}
\end{figure}

A conceptual model that represents the entities of our testbed, including its inputs and outputs -- shaded green and red respectively -- is provided in Figure~\ref{fig:conceptual}. External users are either normal users or attackers (DoS / DDoS). External users generate network flows, which then go to local users. The inputs to \system{} are: (a) the attributes of individual network flows, i.e., \texttt{protocol}, \texttt{port}, and \texttt{packet\_size}, and (b) aggregated flow attributes computed for all the flows passing through \system, i.e., \texttt{num\_flows} and \texttt{num\_unreplied}.  Here, \texttt{protocol} specifies the communication protocol, e.g., \texttt{TCP} or \texttt{UDP}; \texttt{port} specifies the destination port number for the incoming traffic;   \texttt{packet\_size} specifies the size of incoming packets; \texttt{num\_flows} denotes the number of all flows passing through \system; and \texttt{num\_unreplied} indicates the number of flows that are not acknowledged by local users.

\system\ updates its state in response to the network flows passing through it, selecting one of the following: \texttt{Safe}, indicating no signs of an attack; \texttt{Warning} indicating the presence of unusual but not necessarily harmful network flows; and \texttt{Alert} indicating the detection of potentially harmful or malicious traffic. In addition, there are the \texttt{Tending Warning} and \texttt{Tending Alert} states used when \system\ is inclined to transition to \texttt{Warning} or \texttt{Alert}, respectively, but has not conclusively committed  to escalation. As Figure~\ref{fig:conceptual} shows, the inputs and outputs of \system\ are time-series vectors. A time-series vector is a function \hbox{$v: [0..T] \rightarrow D$}, where the interval $[0..T]$ is a \emph{time domain} with duration $T$, and $D$ is the range of the time series.

In a real-world deployment, \system\ has access to all the inputs shown in  Figure~\ref{fig:conceptual}, i.e., the attributes highlighted green,  \emph{except for} the type of the external users: in such a deployment, one does not know whether the  flows are attack flows or normal flows. Therefore, based on raw data from a real-world deployment, one cannot determine whether \system\  accurately identifies attacks and responds appropriately. To verify system behaviours, we need to associate time-series vectors from network flows with the user types generating these flows. Our testbed establishes this association, enabling the learning of state machines that reliably link system behaviours to the external user types -- attackers or normal users -- generating input flows.

\begin{figure}[t]
    \centering
    \includegraphics[width=.85\linewidth]{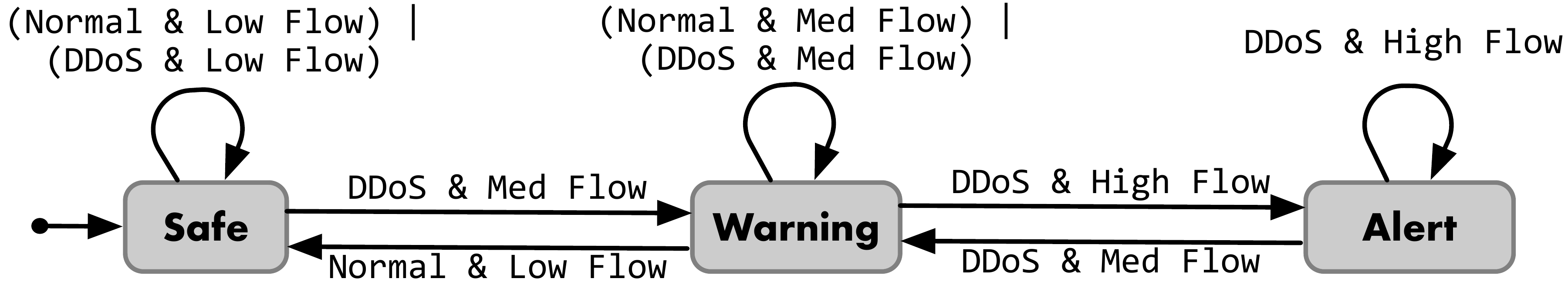}
      \vspace*{-.2cm}
    \caption{A \emph{simplified} example of a state machine learned for \system{} by our approach (\app).}
    \Description{A \emph{simplified} example of a state machine learned for \system{} by our approach (\app).}
    \label{fig:SM}
       \vspace*{-.3cm}
\end{figure}

In the next section, we present \app, our approach for deriving state machines from time-series data.  A \emph{simplified} example state machine generated for \system\ using \app\ is depicted in  Figure~\ref{fig:SM}.  This state machine shows how \system\ changes state in response to receiving a Low, Med(ium) or High number of flows from normal users or DDoS attackers. \app\ uses a systematic process, guided by machine learning,  to derive effective abstractions from  numeric time-series data obtained from \system's inputs and outputs. In Sections~\ref{sec:eval} and~\ref{sec:lessons}, we discuss how these abstractions not only contribute to the conciseness and accuracy of the learned state machines, but also aid RRT engineers  in verifying the system-level requirements of \system\ and discovering unknown behaviours such as those related to the \texttt{Tending Warning} and \texttt{Tending Alert} states (not included in Figure~\ref{fig:SM} due to space). In particular, our state machines enable engineers to (a) identify the ranges for the number of normal and attack flows and (b) refine the requirements of  \system{} regarding how it should react to normal and attack scenarios with different flow sizes. 

%% file: Sections/approach.tex
\section{ML-Enhanced Automata Learning}
\label{sec:approach}
Algorithm~\ref{alg:approach} shows our approach for ML-enhanced automata learning (MELA). The input to the algorithm is a system, \texttt{S}, denoting the system under learning (SUL).
\app\ assumes that \texttt{S}
 accepts time-series data as input and generates time-series data as output. Examples of such systems include cyber-physical systems (CPS) and network systems~\cite{JodatNSS23, TOSEM}.  Our approach treats \texttt{S} as a black box and does not make any assumptions about its internals. In addition to \texttt{S}, \app\ requires four input parameters. These parameters, as we will discuss later, are used in the routines responsible for generating traces and abstracting numeric values in time-series data (lines 7--8 of Algorithm~\ref{alg:approach}). The output of \app\ is a state machine derived based on the time-series data obtained from \texttt{S}.

\begin{algorithm}[t]
\caption{ML-enhanced automata learning (\app) for systems with time-series inputs and outputs.}
\label{alg:approach}
\begin{flushleft}
\small
\textbf{Input} \texttt{S}: System under learning\\
\textbf{Param} $\delta$: Sampling rate \\
\textbf{Param} \texttt{Max$\_$Depth}: Maximum depth of decision trees \\
\textbf{Param} \texttt{Sup$\_$Th}: Support threshold for range abstraction \\
\textbf{Param} \texttt{Purity$\_$Th}: Confidence threshold for range abstraction \\
\textbf{Output} \texttt{Aut}: An automaton abstracting the behaviour of \texttt{S}\\
\end{flushleft}
\begin{algorithmic}[1]
\small
\State \texttt{TimeSeriesData} = $\emptyset$;
\State  \textbf{do} \Comment{Data Generation Loop}
\State  \hspace{0.2cm} \texttt{Input = }\Call{GenerateInput}{\texttt{S}}; 
\State   \hspace{0.2cm} \texttt{Output = }\Call{Execute}{\texttt{Input}, \texttt{S}};
\State \hspace{0.2cm} \texttt{TimeSeriesData} = \texttt{TimeSeriesData} $\cup$ (\texttt{Input}$\cdot$\texttt{Output});
\State \textbf{until} \texttt{(state coverage is not improving)}
\State  \texttt{Traces = }\Call{CreateTraces}{\texttt{TimeSeriesData}, $\delta$};
\State  \texttt{Traces$'$ = }\Call{AbstractTraces}{\texttt{Traces}, \texttt{Max$\_$Depth}, \texttt{Sup$\_$Th}, \texttt{Conf$\_$Th}};
\State  \texttt{Aut = }\Call{LearnAutomata}{\texttt{Traces$'$}}; 
\State \Return \texttt{Aut};
\end{algorithmic}
\end{algorithm}

\begin{figure}[t]
    \centering
    \includegraphics[width=.85\linewidth]{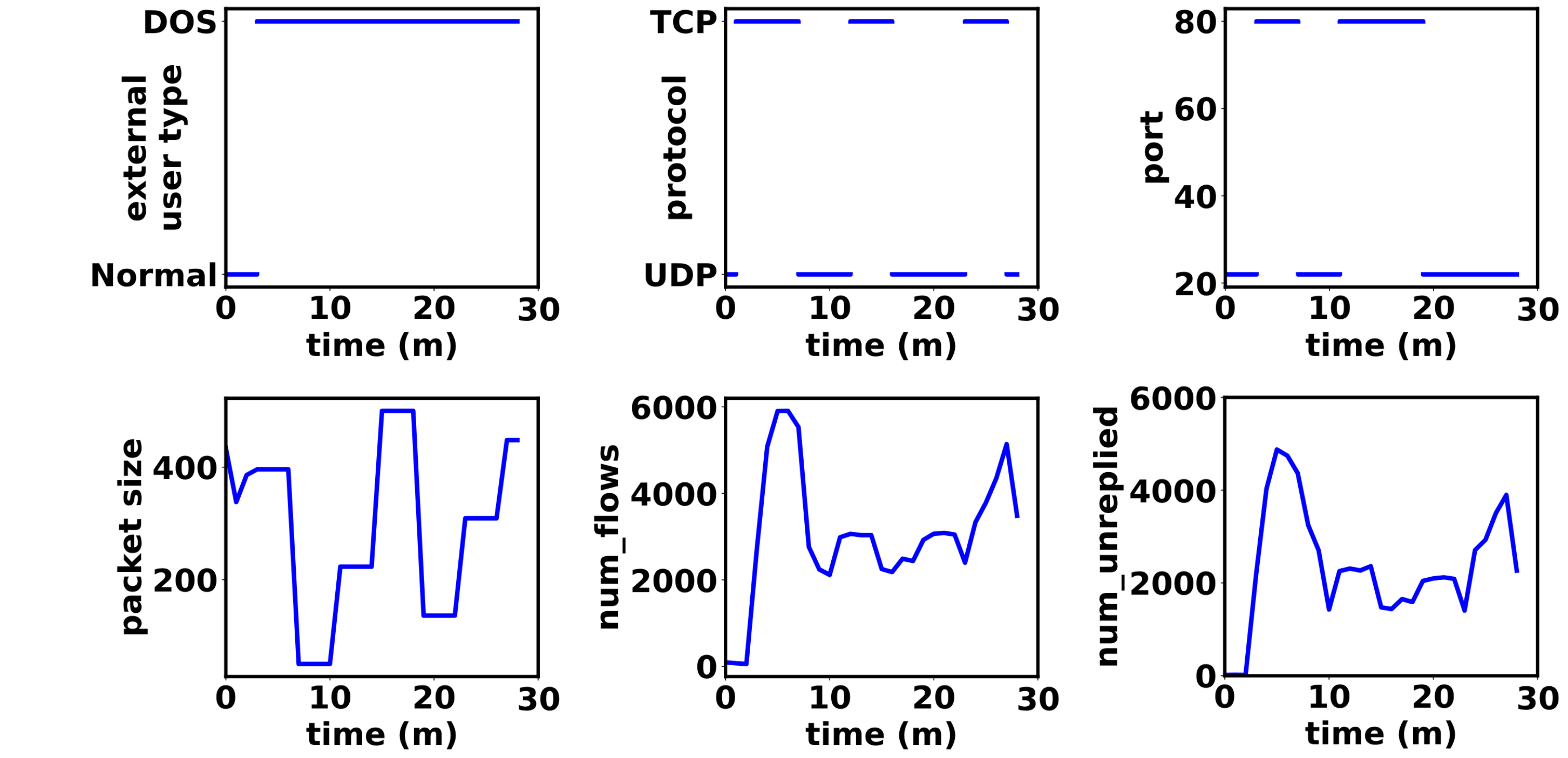}
    \vspace*{-.2cm}
    \caption{A test input consisting of time-series vectors corresponding to six inputs of \system.}
    \Description{A test input consisting of time-series vectors corresponding to six inputs of \system.}
    \label{fig:inputs}
       \vspace*{-.3cm}
\end{figure}

Following the formalization of signal-based test inputs and outputs for CPS~\cite{GaaloulMNBW20,MatinnejadNB17,MatinnejadNBB14}, we denote a test input for \texttt{S} as  $\inputs=(\inputsignal_1, \inputsignal_2 \ldots \inputsignal_m)$ and a test output for \texttt{S}  as $\outputs= (\outputsignal_1, \outputsignal_2 \ldots \outputsignal_n)$ where $m$ is the number of system inputs, $n$ is the number of system outputs, and each $\inputsignal_j$ and each  $\outputsignal_j$  is a time-series vector for some input and some output of \texttt{S}, respectively.  Given a time-series vector $v: [0..T] \rightarrow D$, the time-series  range $D$ can be either discrete, i.e., $D \subseteq \mathbb{N}$, or continuous, i.e., $D \subseteq \real$.  Discrete time-series ranges can be enumerate such as $D = \{\texttt{UDP}, \texttt{TCP}\}$, or numeric such as $D = \{1, \ldots, 6000\}$. As discussed in Section~\ref{sec:intro}, automata learning algorithms are not able to abstract and generalize numeric data ranges, whether continuous or discrete.  Figure~\ref{fig:inputs} illustrates time-series vectors over the time domain $[0..30\,\text{min}]$ for the inputs of \system{}, i.e., the attributes shaded green in Figure~\ref{fig:conceptual}: external user type, protocol, port, packet size, and \system{}'s number of flows and number of unreplied requests. The external user type and protocol have an enumerated range, while the other four inputs are numeric.

We assume that among the outputs of system \texttt{S}, there is one output that captures the system state and has a discrete, enumerable range. For example, as shown in Figure~\ref{fig:conceptual},  \system\ has a specific system-state output. A system-state output is often present in CPS and network systems, which require continuous monitoring and control~\cite{JodatNSS23,GaaloulMNBW20,AtaiefardMHW22}. This output offers feedback on the system's operational status and helps with decision making and control.

\textbf{Data Generation.} Algorithm~\ref{alg:approach} begins with a data generation loop (lines 2–6), where it iteratively generates test inputs for \texttt{S} and executes system \texttt{S} to produce test outputs. The purpose of the data generation loop is to produce time-series data to be used for automata learning. In this loop, the test inputs  are randomly generated using existing parameterized time-series data generation techniques~\cite{signals19,pareto18}.  To ensure that the learned automata effectively capture the behaviours of SUL, we need to generate test inputs that exercise SUL for different scenarios and yield test outputs that adequately capture SUL's behaviours. To increase the adequacy of the generated data,  we employ established black-box test coverage criteria for software testing based on system-state coverage~\cite{offuttTesting}. Specifically, the data generation loop of Algorithm~\ref{alg:approach} terminates when there is no further improvement in state coverage. This occurs either when the generated outputs cover all system states or when, after several consecutive iterations, the outputs do not cover any new states, indicating that our test generation is unlikely to yield further improvements in state coverage.

\textbf{Trace Creation.} Algorithm~\ref{alg:approach} uses the \textsc{CreateTraces} routine (line 7) to convert time-series data vectors into traces to be used for autamata learning. Each time-series data vector  $v: [0..T] \rightarrow D$ is converted into a sequence $v^0, v^1, \ldots, v^k$ of values using the sampling rate parameter $\delta$, which is an input to Algorithm~\ref{alg:approach}. Specifically, $v^0 = v(0)$, $v^1 = v(\delta), v^2 = v(2 \cdot \delta), \ldots, v^k = v(k \cdot \delta)$, with $k \cdot \delta = T$.

Let $\inputs = (\inputsignal_1, \inputsignal_2 \ldots \inputsignal_m)$ be a test input of \texttt{S}, and $\outputs = (\outputsignal_1, \outputsignal_2 \ldots \outputsignal_n)$ be a test output of \texttt{S}. An input/output trace corresponding to each pair of test input and output of \texttt{S} is defined as follows: 
{\small$$(\inputsignal^0_1, \ldots, \inputsignal^0_m, \outputsignal^0_1,  \ldots, \outputsignal^0_n), 
(\inputsignal^0_1, \ldots, \inputsignal^0_m, \inputsignal^1_1, \ldots, \inputsignal^1_m, \outputsignal^1_1,  \ldots, \outputsignal^1_n), 
\ldots,$$ 
$$(\inputsignal^0_1, \ldots, \inputsignal^0_m, \ldots, \inputsignal^k_1, \ldots, \inputsignal^k_m, \outputsignal^k_1,  \ldots, \outputsignal^k_n)$$}

where $k$ is the number of steps with time-step size $\delta$ in time domain $[0..T]$, and for every $l$, $\inputsignal^j_l$ is the $j$th sampled value from input vector $\inputsignal_l$, and $\outputsignal^j_l$ is the $j$th sampled value from output vector~$\outputsignal_l$.

\begin{figure}[t]
    \centering
    \includegraphics[width=.85\linewidth]{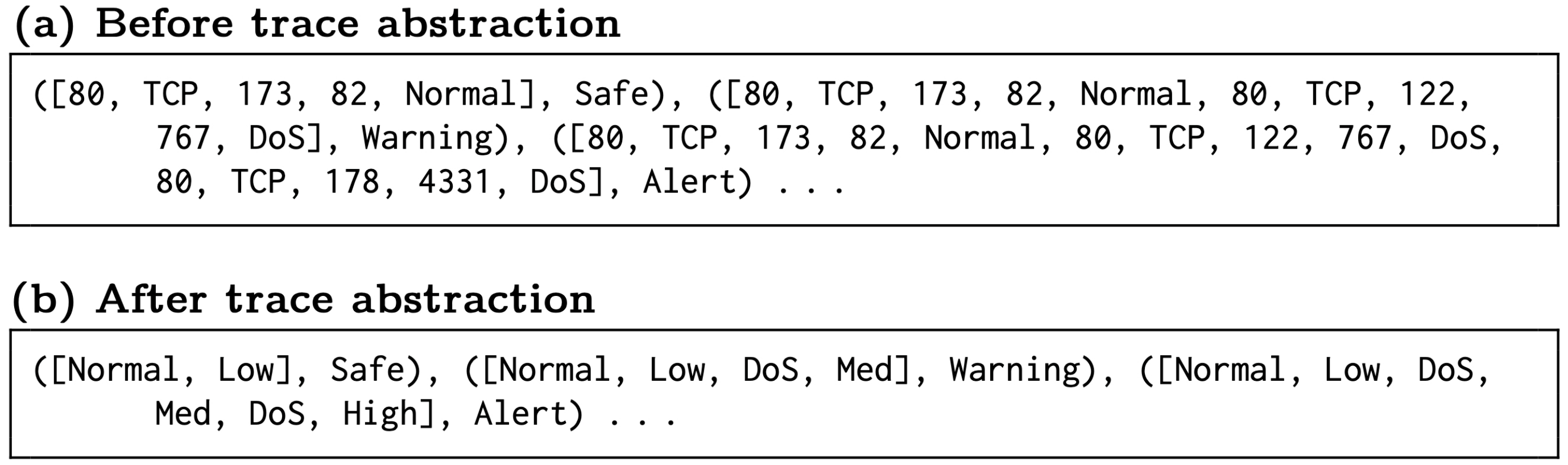}
\vspace*{-.2cm}
    \caption{Traces for \system: (a) an example of an actual trace and (b) the same trace after trace abstraction.}
    \Description{Traces for \system: (a) an example of an actual trace and (b) the same trace after trace abstraction.}
    \label{fig:trace_abstraction}
       \vspace*{-.3cm}
\end{figure}

For example, Figure~\ref{fig:trace_abstraction}(a) shows a small excerpt from a trace generated for \system{}. The values enclosed within ``['' and ``]'' are, respectively, sampled from the following test input vectors of \system: \texttt{port}, \texttt{protocol}, \texttt{packet\,\,size}, \texttt{num\_flows} and \linebreak \texttt{External user type}. The last value in each tuple is sampled from the test output vector of \system: \texttt{State}.

\textbf{Trace Abstraction.} Automata learning approaches assume that traces consist of abstract values only~\cite{muvskardin2022aalpy}. Hence, raw numerical values should be replaced by categorical or interval-based representations before applying automata learning. To obtain traces consisting of abstract values, we use the \textsc{AbstractTraces} routine  (line~8), which is a novel contribution of our work. This routine uses statistical machine learning to refine traces consisting of raw numerical values into a more abstract form. The \textsc{AbstractTraces} routine consists of two steps:  \emph{First,} among all the inputs and outputs of \texttt{S}, we select those that are non-redundant and most correlated with the system state. \emph{Second,}  we abstract raw, numeric ranges of the inputs and outputs of \texttt{S} into discrete categories.  Below, we describe these two steps; we refer to the first step as \emph{variable selection} and to the second step as \emph{range abstraction}.

\emph{Variable selection.} We use the information gain~\cite{mlbook} to calculate the importance score of each input and each output of \texttt{S} with respect to predicting the system state. To do so, we create a table using the data from our traces such that each column lists the values appearing in the traces for an individual input or output, and the last column lists the corresponding values for the system state. We then compute the information gain of all the columns in this table with respect to the last column. The information gain quantifies the predictive relevance of each column, which represents a system input or output, towards the system state. We rank these columns based on their importance score and eliminate the lowest-ranked ones (i.e., select the highest-ranked ones). We then refine the traces obtained from the trace creation routine (line~7 of Algorithm~\ref{alg:approach}) by removing the values related to the eliminated inputs and outputs. 

For example, by applying variable selection to \system, we 
remove the variables \texttt{num\_unreplied}, \texttt{port}, \texttt{protocol}, and \texttt{packet size}, and retain only \texttt{num\_flows} and \texttt{External User type}. This modification follows the computation of importance scores, where \texttt{num\_flows} and \texttt{External User type} are the top-ranked with a large distance from the other variables in terms of importance.

\emph{Range abstraction.} We use decision-tree learners to abstract the ranges of numeric inputs and outputs included in our traces after the variable selection step.  For each numeric input or output variable $u$, we create a two-column table where the first column is the values of $u$ appearing in the traces, and the second column is the corresponding values for the system state.  We then develop a decision tree with inputs as the values of $u$  and outputs as the values of the system state which is a categorical variable.  The tree's construction is controlled by a stopping criterion specified by the input parameter \texttt{Max$\_$Depth} of  Algorithm~\ref{alg:approach}. This parameter determines the maximum depth to which the tree can grow. Each tree leaf represents the following information: (1)~the count of samples that are clustered in that leaf (support), and (2) the purity of the leaf (confidence), which indicates the homogeneity of the samples within the leaf. A higher purity  indicates a greater concentration of one class; in our context, each class  corresponds to a system state. Every leaf is linked to its immediate parent node through a condition such as  $u < c$, where $u$ is the variable and $c$ is a constant in the range of $u$. Among all the tree leaves, we select the ones that have a number of samples and a purity percentage, respectively, higher than the \texttt{Sup$\_$Th} and  \texttt{Purity$\_$Th} thresholds, which are input parameters of Algorithm~\ref{alg:approach}. We consider the conditions $u < c$ linking the selected leaves to their immediate parent nodes.  Let $c_1, \ldots, c_k$ denote the constants in ascending order that appear in these conditions. We partition the range of $u$ into the following intervals: $[0, c_1), [c_1, c_2) \ldots [c_k, \infty)$. Then in our traces, we replace the numeric values of $u$ with \hbox{the categories 
representing these intervals.} 

\begin{figure}[t]
    \centering
    \includegraphics[width=.85\linewidth]{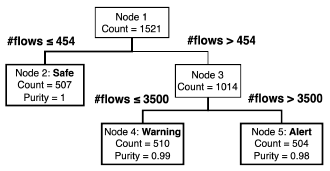}
     \vspace*{-.2cm}
            \caption{Illustrating how a decision tree is used to abstract the numeric range of the \texttt{num\_flows} input attribute. The numeric range of \texttt{num\_flows} is abstracted to the enumerated range [Low, Med, High] such that ``Low''=$[0,454)$, ``Med''=$[454,3500)$, and ``High''=$[3500, \infty)$.}
            \Description{Illustrating how a decision tree is used to abstract the numeric range of the \texttt{num\_flows} input attribute. The numeric range of \texttt{num\_flows} is abstracted to the enumerated range [Low, Med, High] such that ``Low''=$[0,454)$, ``Med''=$[454,3500)$, and ``High''=$[3500, \infty)$.}
    \label{fig:DT}
        \vspace*{-.3cm}
\end{figure}

For example,  Figure~\ref{fig:DT} shows a decision tree used to abstract the numeric \texttt{num\_flows} variable. As shown in Figure~\ref{fig:trace_abstraction}(b), our traces for \system\ include tuples relating values of \texttt{num\_flows} and other input variables of \system\ to system-state values. The decision tree in Figure \ref{fig:DT}  determines, based on the  \texttt{num\_flows} and \texttt{state} values extracted from traces, how well \texttt{num\_flows} predicts the system state.  For this example, we assume that \texttt{Max$\_$Depth} is set to 3, and the \texttt{Sup$\_$Th}  and  \texttt{Purity$\_$Th} thresholds are set to $20$\% of the total data and to $70$\%, respectively. 
In Figure~\ref{fig:DT}, we show the generated tree leaves and their respective number of samples and purity level. We select all three leaves in the figure, i.e., Node~2, Node~4 and Node~5, since for all these leaves, both the sample count and purity are greater than their respective thresholds. Based on the conditions linking these leaf nodes to their parent nodes, we partition the range of \texttt{num\_flows} into the following intervals: $[0, 454)$, $[454, 3500)$, $[3500, \infty)$. The number of intervals is not fixed a priori and depends on the decision tree. For \texttt{num\_flows}, as we have three intervals, we designate them as ``Low'', ``Med'', and ``High'', respectively, for better readability. Specifically: ``Low''=$[0,454)$, ``Med''=$[454,3500)$, and ``High''=$[3500, \infty)$.
After performing variable selection and range abstraction for \system, 
we convert the trace in Figure~\ref{fig:trace_abstraction}(a) to that in Figure~\ref{fig:trace_abstraction}(b).

\textbf{Automata Learning.} Algorithm~\ref{alg:approach} uses the \textsc{LearnAutomata} routine to build an automaton (line~9) from abstract traces. We realize this routine using the well-known Regular Positive Negative Inference (RPNI)~\cite{de2010grammatical, cano2010inferring, muskardin2022active} algorithm which is for passive learning. While the original RPNI algorithm has been designed to generate DFAs, there are variants of RPNI that can learn Mealy and Moore machines~\cite{muvskardin2022aalpy}, which are more expressive models of computation wherein outputs depend not only on the current state but also on inputs. For our case-study system, \system, we choose to generate \emph{Moore machines}, as this system's traces include both inputs and outputs.  Furthermore, Moore machines map outputs directly to states, making them a suitable representation for \system, as the output of \system{} reflects the system's state. In the rest of this paper, depending on the context, we interchangeably refer to the outputs of our approach  -- Moore machines -- as either state machines or automata.

For example, Figure \ref{fig:SM} shows a simplified state machine learned for \system, illustrating how it transitions between different states -- \texttt{Safe}, \texttt{Warning}, and \texttt{Alert} -- based on the number of flows -- Low, Med(ium), or High -- and different external user types -- Normal or DDoS. Briefly, \system{} starts in the \texttt{Safe} state and maintains this state while there is no attack, i.e., Normal users sending flows, or when an attack involves low numbers of flows. The system transitions to the \texttt{Warning} state under a medium-flow DoS attack, escalating to the \texttt{Alert} state under a high-flow DoS attack.

%% file: Sections/eval.tex
\section{Evaluation}
\label{sec:eval}

We evaluate \app\ by (1) assessing the complexity and accuracy of the generated state machines in representing the behaviours of our case-study system (\system), and (2) investigating how the learned state machines help with verifying this system against its requirements and exploring unknown behaviours. Specifically, our evaluation aims to answer the following research questions (RQs):

\textbf{RQ1 (Complexity and Conformance).} \emph{How effective is the trace abstraction component of \app\ in reducing the complexity of the generated state machines while maintaining a high-level of accuracy?} 
RQ1 assesses how the trace abstraction component of \app, specifically line~8 of Algorithm~\ref{alg:approach}, impacts the complexity and accuracy of the generated state machines. The primary goal of the \app\ is to generate state machines that are understandable, abstract, yet highly accurate, ensuring high conformance to SUL. As a baseline for comparison, we consider an approach that incorporates a trace abstraction component similar to \app, but instead of relying on statistical machine learning, it uses manually defined abstractions based on expert judgment. We refer to this baseline as \textsc{Manual}. Note that in \textsc{Manual}, experts manually abstract numeric value ranges but do \emph{not} manually create states and transitions. States and transitions are still inferred automatically through automata learning, similar to \app.

\textbf{RQ2 (Verification).} \emph{Do the state machines learned using \app\ help determine whether the system meets its requirements and explore its unknown behaviours?} In collaboration with  RRT, we elaborate the system-level requirements of \system\ into detailed temporal properties. As discussed in Section~\ref{sec:testbed},  the behaviours of \system{} is not fully known, particularly for the circumstances under which the system may enter the \texttt{Tending Warning} and~\texttt{Tending Alert} states. To explore unknown system behaviours, in addition to temporal properties, we develop \emph{temporal queries} which are temporal properties with placeholders~\cite{Chan00}.  Temporal queries yield predicates such that, when these predicates replace the placeholder in the query, they form a property that holds over the state machine. We report on the evaluation of our temporal properties and queries against the state machines learned in RQ1.

\subsection{Testbed Implementation}
\label{subsec: implementation}

We generate the time-series input data for \app\ using the testbed presented in Figure~\ref{fig:fig_1}. Our testbed consists of three virtual machines (VMs) created using VirtualBox~\cite{virtualbox} and  deployed on three separate laptop computers. The left VM, i.e., VM-left,  simulates local users, the centre VM, i.e., VM-centre,  hosts the \system{} from RRT, and the third VM, i.e., VM-right,  simulates external users. Both attackers and normal users are simulated using VM-right, while VM-left is designed to capture potential vulnerabilities in local users. VM-right generates two types of data flows: normal traffic and DoS/DDoS packets, using the \emph{hping3} tool~\cite{hping3} -- a network tool designed for testing firewall rules and network performance. After passing through \system~(VM-centre), these flows are directed to VM-left, running Metasploitable~\cite{metasploit} -- an intentionally vulnerable Linux VM for security and penetration testing. VM-left and VM-right are connected to \system{} using two unmanaged NETGEAR GS308v3 Gigabit Ethernet switches. The time-series values for all input and output variables of the system, except for the values of the \texttt{external user type} variable, are gathered from VM-centre. The values for \texttt{external user type},  which are not accessible through VM-centre, are collected from VM-right.

\subsection{RQ1: Complexity and Conformance}
We discuss the experiment design and the results obtained for RQ1.
\subsubsection{Baseline}
\label{subsec:baseline}
We compare \app\ with the \textsc{Manual} baseline,  introduced earlier. This baseline obtains traces similarly to \app, using the same data generation loop and trace creation steps, i.e., lines~1--7 of  Algorithm~\ref{alg:approach}. However, \textsc{Manual} differs from \app\ on line 8: Although \textsc{Manual} uses a two-step data abstraction process, similar to that of \app, which encompasses variable selection and range abstraction, it relies on expert knowledge rather than machine learning. Specifically, for variable selection, a domain expert manually selects the most relevant input and output variables of \system\ for identifying DoS and DDoS attacks. In the range abstraction step, the expert abstracts the numeric ranges by partitioning them into enumerated categories of his choice.

\subsubsection{Experiment Design}
\label{sec:expdesign}
To compare \app\ with \textsc{Manual}, we use identical sets of traces as inputs by applying the data generation and trace creation steps of Algorithm~\ref{alg:approach} over the testbed described in Section~\ref{subsec: implementation}. We refer to the generated trace sets, which are used to learn automata, as \emph{learning sets}. We use time-series vectors with a time domain $[0..256min]$ for the input generation routine, i.e., line~3 of Algorithm~\ref{alg:approach}, and the sampling rate $\delta=10sec$ to convert the time-series data vectors into traces by the trace creation routine, i.e., line~7 in Algorithm~\ref{alg:approach}. The choice of sampling rate is based on the minimum refresh rate that our testbed supports. A long duration in the time domain was recommended by the domain expert  to enable us to obtain traces achieving higher state coverage.

As discussed in Section~\ref{sec:testbed}, \system's inputs consist of normal traffic combined with either DoS or DDoS attacks.  We do not alternate between DoS and DDoS attacks during an individual testing campaign, as each type of attack requires a different experimental setup. Hence, we develop distinct learning sets for DoS and DDoS attacks, resulting in separate state machines capturing the behaviours of \system\ for each attack type. To account for randomness in the generation of the learning sets, we rerun the data generation loop of Algorithm~\ref{alg:approach} five times for DoS and five times for DDoS, obtaining  five different learning sets for each attack type. For DoS, three of these learning sets covered only three system states (\texttt{Safe}, \texttt{Warning}, and \texttt{Alert} in Figure~\ref{fig:conceptual}), while the other two sets covered all five system states in Figure~\ref{fig:conceptual}. For DDoS,  all generated learning sets could only cover three out of the five system states, i.e., \texttt{Safe}, \texttt{Warning}, and \texttt{Alert}. This inability to achieve full state coverage led to further investigation, as we will explain in Section~\ref{sec:rq2}. This investigation revealed, as confirmed by our domain expert, that the two states, \texttt{Tending Warning} and~\texttt{Tending Alert}, are unreachable under DDoS attacks.

In view of the above, we use the following learning sets for our experiments: \texttt{DoS3}, the union of the three learning sets for DoS with three-state coverage; \texttt{DoS5}, the union of the two learning sets for DoS with five-state coverage; and \texttt{DDoS}, the union of the five learning sets for DDoS. We note that one could obtain a single learning set for DoS by combining all five learning sets. 
However, as indicated in our online material~\cite{neayoughi2024mela}, the state machines obtained by unioning all five sets and those obtained by combining the two five-state coverage sets are identical.  Hence, in the paper, we present the results for two separate DoS learning sets -- one for three-state and another for five-state coverage -- to facilitate direct comparison with the DDoS results, which only achieve three-state coverage. Table~\ref{tab:param}(a) shows the average length of traces in \texttt{DoS3}, \texttt{DoS5}, and \texttt{DDoS}, as well as the total execution time (in minutes) required to generate the traces in these learning sets.

\begin{table}[t]
\centering
\caption{Parameters for our experiments: (a)~parameters of the learning sets used by \app\ and the baseline; (b)~parameters of the trace abstraction step of \app; and (c)~information about trace abstraction in the \textsc{Manual} baseline.}\label{tab:param}
\vspace*{-.2cm}
\resizebox{\columnwidth}{!}{%
\begin{tabular}{@{}llcc@{}}
\toprule
\multicolumn{4}{c}{\cellcolor{gray!40}\textbf{a. Trace Generation for \app and \textsc{Manual} }} \\ \midrule 
& \textbf{Learning Set}  & \textbf{Avg Trace Length} & \textbf{Execution Time (m)} \\ \midrule
& \texttt{DoS3}: 3-state coverage & 1530  & 870  \\
& \texttt{DoS5}: 5-state coverage & 1402  & 474  \\
& \texttt{DDoS}: 3-state coverage & 1523  & 1420  \\ \midrule
\multicolumn{4}{c}{\cellcolor{gray!40}\textbf{b. Trace Abstraction for \app}}                                         \\ \midrule
\multicolumn{2}{l}{\textbf{Variables Selected by Information Gain}}  &  \multicolumn{2}{l}{\textbf{Range Abstraction by Decision Tree}} 
\\ \midrule
\multicolumn{2}{l}{\texttt{1st}: \texttt{num\_flows}} &  \texttt{Max$\_$Depth}: 3 \\
\multicolumn{2}{l}{\texttt{2nd}: \texttt{num\_unreplied}} &
  \texttt{Purity$\_$Th}: 0.7\\
\multicolumn{2}{l}{\texttt{Top-2}: \texttt{num\_flows} and \texttt{num\_unreplied}} & \texttt{Sup$\_$Th}: $0.2\times n; n$ is total data\\ \midrule
\multicolumn{4}{c}{\cellcolor{gray!40}\textbf{c. Trace Abstraction for the \textsc{Manual} baseline}}                                       \\ \midrule
\multicolumn{2}{l}{\textbf{Variables Selected by Domain Expert}} & \multicolumn{2}{l}{\textbf{Range Abstraction by Domain Expert}} \\ \midrule
& & \multicolumn{2}{l}{Let $d$ be the max range of the numeric variable: } \\
\multicolumn{2}{l}{\texttt{1st}: \texttt{num\_flows}} & \multicolumn{2}{l}{\ \ \ \texttt{Low}: $[0...0.33\times d]$}\\
\multicolumn{2}{l}{\texttt{2nd}: \texttt{num\_unreplied}} & \multicolumn{2}{l}{\ \ \ \texttt{Med}: $[0.33\times d...0.66\times d]$}  \\
\multicolumn{2}{l}{\texttt{Top-2}: \texttt{num\_flows} and \texttt{num\_unreplied}} & \multicolumn{2}{l}{\ \ \ \texttt{High}:  $[0.66\times d...d]$}  \\ \bottomrule
\end{tabular}
}

\end{table}

For the trace abstraction step of \app, i.e., line~8 of Algorithm~\ref{alg:approach}, we perform variable selection and range abstraction. For variable selection,  we rank \system's input variables, i.e., variables shaded green in Figure~\ref{fig:conceptual}, based on their information gain relative to the system state. In our experiments, there is a significant information-gain gap between the second- and third-ranked variables. Therefore, we consider three alternative cases involving the selection of the two top-ranked variables of the system: (1)~selecting the variable with the highest information gain, i.e., \texttt{1st}: \texttt{num\_flows};  (2)~selecting the variable with the second-highest information gain, i.e., \texttt{2nd}: \texttt{num\_unreplied}; and (3)~selecting the top two variables with the highest information gains, i.e., \texttt{Top-2}: \texttt{num\_flows} and \texttt{num\_unreplied}. For range abstraction, we use a decision tree to convert the numeric ranges of the \system's variables into enumerated ranges. We set the tree's maximum-depth parameter in Algorithm~\ref{alg:approach} to three (i.e., \texttt{Max$\_$Depth} = 3); this helps avoid overfitting and prevents our trees from generating too many leaves, which may result in partitioning numeric ranges into several fine-grained intervals and having overly detailed enumerated ranges. Further, we set the \texttt{Sup$\_$Th} parameter to $20$\% of the total data count used for building the decision tree, and the \texttt{Purity$\_$Th} parameter to $70$\%. This ensures that the tree leaves selected for defining partitions contain a sufficient number of elements corresponding to a specific system state. The details related to the trace abstraction component of  \app\  for our experiments are presented in Table~\ref{tab:param}(b).

For the \textsc{Manual} baseline, we use the domain expert's judgment for variable selection and range abstraction. Specifically, for variable selection, we requested the domain expert to select the two variables that, in his opinion, most significantly impact the state of \system.  The variables selected by the expert matched those identified by our approach. For range abstraction, the expert suggested dividing the ranges of each numeric variable into three equal intervals of \texttt{Low}, \texttt{Med} and \texttt{High}. The information related to the trace abstraction for the \textsc{Manual} baseline is shown in Table~\ref{tab:param}(c). 
Finally, for automata learning, both \app\ and \textsc{Manual} use the RPNI passive learning algorithm from AALpy~\cite{muvskardin2022aalpy} -- a Python library that provides a range of advanced automata learning algorithms.

\subsubsection{Metrics}
\label{subsec:metric}
To assess the complexity of the generated state machines, we report the number of states, the number of transitions, and the size of the input alphabet. These three size-based metrics are commonly used in the literature to evaluate the complexity of state machines~\cite{Hall11}. To measure the accuracy of the learned models, we follow the established practice in the automata learning literature~\cite{muskardin2022active} and compare the models against the ground truth, i.e., traces generated by the \system's testbed in our context.  We note that the learned automata in passive learning are only as good as the traces in the learning sets. Since the learning sets might be incomplete, there could be an accuracy gap between the behaviours of the learned automata and those of the actual system. 

To measure the accuracy of \app\ and the baseline, we generate five randomly generated sets of traces and determine the percentage of these traces that the automaton generated by each technique can accept.  We refer to these sets as \emph{test sets}. We use the data generation loop of Algorithm~\ref{alg:approach} to create these test sets. To ensure that we cover a range of trace lengths, we generate five test sets referred to as  \emph{very small}, \emph{small}, \emph{medium}, \emph{large}, and \emph{very large}. Each test set has $100$ traces, leading to a total of $500$ traces for assessing the accuracy of the learned automata.  The average trace lengths in these test sets are as follows: very small at $70$, small at $138$, medium at $207$, large at $276$, and very large at $345$. The lengths of the traces in the very small, small, medium, large, and very large test sets are chosen to be approximately 10\%, 20\%, 30\%, 40\%, and 50\% of the trace length in the learning set, respectively.  It took approximately $600$ hours to generate the traces in these test sets using our testbed discussed in Section~\ref{subsec: implementation}. Given the high cost of trace generation, we decided to cap the max length of traces in test sets at $50$\% of the size of traces in our learning sets. 

\subsubsection{Results}
\begin{figure}[t]
     \centering
    \includegraphics[width=.85\linewidth]{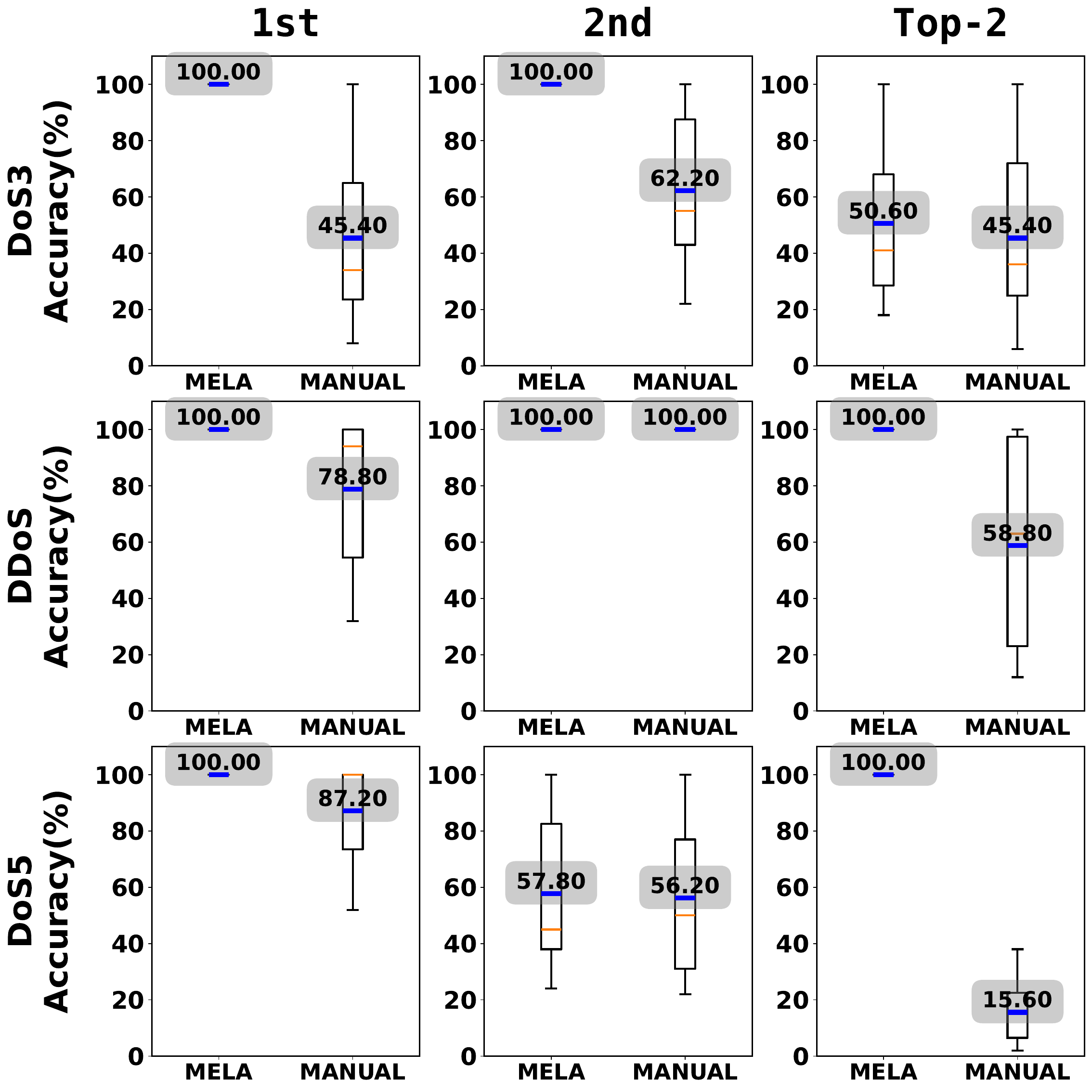}
        \vspace*{-.2cm}
     \caption{Comparing the accuracy of the state machines learned by \app\ versus by the \textsc{Manual} baseline for  different learning sets (\texttt{DoS3}, \texttt{DoS5} and \texttt{DDoS}), and different configurations (\texttt{1st}, \texttt{2nd}, and \texttt{Top-2} as defined in Table~\ref{tab:param}).}
     \Description{Comparing the accuracy of the state machines learned by \app\ versus by the \textsc{Manual} baseline for  different learning sets (\texttt{DoS3}, \texttt{DoS5} and \texttt{DDoS}), and different configurations (\texttt{1st}, \texttt{2nd}, and \texttt{Top-2} as defined in Table~\ref{tab:param}).}
     \label{fig:boxplot}
     \vspace*{-.3cm}
 \end{figure}

To answer RQ1, we apply \app\ and the \textsc{Manual} baseline to the learning sets \texttt{DoS3}, \texttt{DoS5}, and \texttt{DDoS}, and subsequently compare the results using the complexity and accuracy metrics discussed in Section~\ref{subsec:metric}.
 Table~\ref{table:tbl_1} reports the number of states and transitions and the alphabet size for the generated automata by  \app\ and  \textsc{Manual}.  Specifically, we obtain 18 automata by applying the two approaches to the three different learning sets and considering three different abstraction options of \texttt{1st}, \texttt{2nd}, and \texttt{Top-2}. Figure~\ref{fig:boxplot} shows the accuracy results for the learned automata by \app\ and \textsc{Manual}. The accuracy values are computed by applying each of the $18$ learned automata to the $500$ test traces described in Section~\ref{subsec:metric}. The plots in the figure show the accuracy distribution of each \hbox{learned automaton with respect to these test traces.}

\begin{table}[t]
\centering
\caption{Comparing the number of states, number of transitions, and the alphabet size for the state machines learned by \app\ versus by the \textsc{Manual} baseline.}\label{table:tbl_1}
\vspace*{-.2cm}
\resizebox{\columnwidth}{!}{%
\begin{tabular}{|c|c|c|c|c|c|c|c|c|c|}
\hline
\multirow{2}{*}{\textbf{Learning set}} & \multirow{2}{*}{\textbf{Configuration}} & \multicolumn{3}{c|}{\textbf{\app}} & \multicolumn{3}{c|}{\textbf{\textsc{Manual}}}  \\
\cline{3-8}
 & & \textbf{\# States} & \textbf{\# Transitions} & \textbf{|Alphabet|} & \textbf{\# States} & \textbf{\# Transitions} & \textbf{|Alphabet|} \\
\hline
\multirow{3}{*}{\textbf{\texttt{DoS3}}} & \texttt{Top-2} & 3 & 11 & 7 & 23 & 65 & 8 \\
\cline{2-8}
 & \texttt{1st} & 3 & 9 & 5 & 290 & 325 & 5\\
\cline{2-8}
 & \texttt{2nd} & 19 & 51 & 5 & 29 & 68 & 5\\
\hline
\multirow{3}{*}{\textbf{\texttt{DDoS}}} & \texttt{Top-2} & 3 & 11 & 7 & 20 & 52 & 8 \\
\cline{2-8}
 & \texttt{1st} & 3 & 9 & 5 & 284 & 319 & 5 \\
\cline{2-8}
 & \texttt{2nd} & 17 & 46 & 5 & 24 & 62 & 5 \\
\hline
\multirow{3}{*}{\textbf{\texttt{DoS5}}} & \texttt{Top-2} & 46 & 83 & 8 & 167 & 253 & 9 \\
\cline{2-8}
 & \texttt{1st} & 47 & 78 & 5 & 410 & 499 & 5 \\
\cline{2-8}
 & \texttt{2nd} & 121 & 187 & 5 & 178 & 267 & 5 \\
\hline

\end{tabular}
}
\end{table}

As indicated by Table~\ref{table:tbl_1} and Figure~\ref{fig:boxplot}, all the 18 generated automata by \app\ have fewer states and transitions than the corresponding autoamta generated by \textsc{Manual}. Through the ML-based trace abstraction in \app, we obtain automata that, on average, have $69.61$\% fewer states and $65.41\%$ fewer transitions than automata  derived using expertise-based abstraction.  Further, the alphabet size of the automata generated by \app\ is the same as or smaller than that of the automata generated by \textsc{Manual}. The accuracy results in Figure~\ref{fig:boxplot} show that \app\ not only substantially reduces the size of the learned automata but also results in significantly more accurate automata. On average, the automata learned by \app\ are $28.75\%$ more accurate than those learned by \textsc{Manual}.

\begin{tcolorbox}[breakable,colback=gray!10!white,colframe=black!75!black]
\textbf{RQ1:} Our approach (\app) leads to an average reduction of 67.5\% in the
number of states and transitions of the learned automata, while
improving accuracy by an average of 28\% compared to using expertise-based
abstractions for automata learning.
\end{tcolorbox}

\subsection{RQ2: Verification}
\label{sec:rq2}

To answer RQ2, we use six state machines from RQ1: (a) three developed using the \texttt{DoS5} learning set (the learning set with the highest coverage for DoS attacks), corresponding to the \texttt{1st}, \texttt{2nd}, and \texttt{Top-2} configurations; and (b) three developed using the \texttt{DDoS} learning set for DDoS attacks, corresponding to the same configurations.

\begin{figure}[t]
     \centering
     \captionof{table}{Model-checking results for the temporal properties derived from \system{} requirements $\mathbf{R_1}$ and $\mathbf{R_2}$}
     \Description{Model-checking results for the temporal properties derived from \system{} requirements $\mathbf{R_1}$ and $\mathbf{R_2}$}
    \includegraphics[width=.85\linewidth]{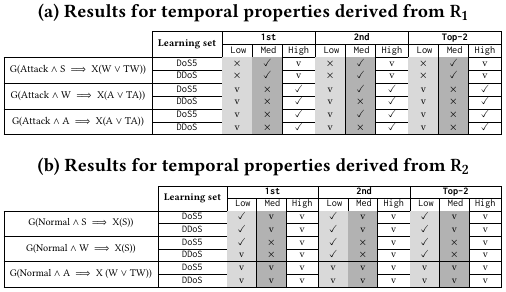}
        \vspace*{-.2cm}
     \label{tbl:main-table}
 \end{figure}

We identify two high-level requirements for \system\ related to its intrusion detection function: \requirement{$R_1$} \emph{When attacks happen,  the system shall change state in a staged manner from safe to warning and from warning to alert}.  \requirement{$R_2$} \emph{When attacks are stopped, the system shall restore in a staged manner its state from alert to warning, and from warning to safe.} In collaboration with RRT, we derived temporal properties from $R_1$ and $R_2$, shown in the leftmost columns of Tables~\ref{tbl:main-table}(a) and~\ref{tbl:main-table}(b), respectively. These properties are expressed as Linear Temporal Logic (LTL) formulas~\cite{Pnueli77,mcbook}, where ``G'' is the globally operator and ``X'' is the next state operator. For succinctness, we use abbreviated names for states: ``S'' for \texttt{Safe}, ``W'' for \texttt{Warning}, ``A'' for \texttt{Alert}, ``TW'' for \texttt{Tending Warning}, and ``TA'' for \texttt{Tending Alert}. For example, the property G,(\text{Attack} $\land$ S $\implies$ X,(W $\lor$ TW)) specifies that after an attack in state S,  the system must transition to W or TW.

The properties in Table~\ref{tbl:main-table}(a), derived from $R_1$, indicate that the system must transition to a higher criticality level in the next state in response to DoS or DDoS attacks. Here, state S represents the lowest criticality state, while TA and A denote the highest. Note that when \system{} is in the TA or A states and an ongoing attack persists, the system shall remain in the TA or A state. In a dual manner, the properties in Table~\ref{tbl:main-table}(b), derived from $R_2$, indicate that the system must transition to a lower criticality level under normal traffic conditions and when \hbox{there is no ongoing attack.}

As discussed in Section~\ref{sec:testbed}, there is uncertainty regarding the system's subsequent state after attack scenarios or under normal traffic conditions for the TW and TA states. Hence, we cannot develop temporal properties having 
 TW and TA as their antecedent similar to those in Tables~\ref{tbl:main-table}(a) and~\ref{tbl:main-table}(b).  Instead of temporal properties, we develop temporal queries with placeholders shown in the leftmost column of Table~\ref{table:qc}.  The queries in Table~\ref{table:qc} enable us to explore the subsequent state of \system{} at the TW and TA states under attack and normal traffic. For example, the temporal query G\,(\text{Attack} $\land$ TA $\implies$ X\,(?)) \hbox{uses a placeholder for the ``X'' operator.}

\subsubsection{Model-checking.}  Tables~\ref{tbl:main-table}(a) and~\ref{tbl:main-table}(b) show the results of verifying the state machines obtained from the \texttt{DoS5} and \texttt{DDoS} learning sets against
the temporal properties for $R_1$ and $R_2$, respectively. For each state machine, a temporal property is marked as pass (\checkmark) if it holds, as fail ($\times$) if it is violated, or as vacuous (v) if the property's antecedent is never met, resulting in vacuous satisfaction~\cite{BeerBER97}. Note that the antecedent refers to the subformulas of the properties in Tables~\ref{tbl:main-table}(a) and~\ref{tbl:main-table}(b) that appear before the $\implies$ operator. When assessing the temporal properties, we observed that the number of attack and normal flows has a significant impact on the results. Therefore, Tables~\ref{tbl:main-table}(a) and~\ref{tbl:main-table}(b) present the results for low, medium, and high numbers of flows separately.

For example, the property G\,(\text{Attack} $\land$ S $\implies$ X\,(W $\lor$ TW))   holds for the state machines learned for both \texttt{DoS5} and \texttt{DDoS} and for the \texttt{1st}, \texttt{2nd}, and \texttt{Top-2} configurations when the number of flows is medium. The property fails when the number of flows is low and is vacuous when the number of flows is high. This indicates that when the system is in its \texttt{Safe} state, requirement $R_1$ is met only when the number of attack flows is medium; a low-flow attack goes undetected, and a high-flow attack never happens. Specifically, the property G\,(\text{Attack} $\land$ S $\implies$ X\,(W $\lor$ TW)) holding vacuously for high-flow traffic implies that the antecedent, \text{Attack} $\land$ S,  never holds when the number of attack flows is high. 

\begin{table}[t]
\centering
\caption{Exploring the behaviours of \system{} at the TW and TA states using temporal queries.}\label{table:qc}
\vspace*{-.3cm}
\resizebox{\columnwidth}{!}{%
\begin{tabular}{|c|c|c|c|c|c|c|c|c|c|c|c|c|}
\cline{2-11}
 \multicolumn{1}{c|}{} & \multirow{2}{*}{\textbf{Learning set}} & \multicolumn{3}{c|}{\textbf{\texttt{1st}}} & \multicolumn{3}{c|}{\textbf{\texttt{2nd}}}  & \multicolumn{3}{c|}{\textbf{\texttt{Top-2}}}\\
\cline{3-11}
\multicolumn{1}{c|}{} & & \ \texttt{Low} & \ \texttt{Med} & \texttt{High} & \ \texttt{Low} & \ \texttt{Med} & \texttt{High} &\ \texttt{Low} & \ \texttt{Med} & \texttt{High}\\
\hline
 \multirow{2}{*}{G(\text{Attack} $\land$ TW $\implies$ X(?))} 
 & \texttt{DoS5} & \cellcolor{gray!30}v & \cellcolor{gray!60}TW $\lor$ TA & A$\lor$ TA  & \cellcolor{gray!30}v & \cellcolor{gray!60}TW & TW $\lor$ TA  & \cellcolor{gray!30}v & \cellcolor{gray!60}TW & TA \\
\cline{2-11}
 & \texttt{DDoS} & \cellcolor{gray!30}v  & \cellcolor{gray!60}v & v & \cellcolor{gray!30}v  & \cellcolor{gray!60}v & v & \cellcolor{gray!30}v & \cellcolor{gray!60}v & v\\
\cline{2-11}
\hline
\multirow{2}{*}{G(\text{Attack} $\land$ TA $\implies$ X(?))} & \texttt{DoS5} & \cellcolor{gray!30}v  & \cellcolor{gray!60}W & A $\lor$ TA  & \cellcolor{gray!30}v & \cellcolor{gray!60}W & A $\lor$ TA  & \cellcolor{gray!30}v & \cellcolor{gray!60}W & A $\lor$ TA \\
\cline{2-11}
 & \texttt{DDoS} & \cellcolor{gray!30}v  & \cellcolor{gray!60}v & v & \cellcolor{gray!30}v  & \cellcolor{gray!60}v & v & \cellcolor{gray!30}v & \cellcolor{gray!60}v & v\\
\cline{2-11}
\hline
\multirow{2}{*}{G(\text{Normal} $\land$ TW $\implies$ X(?))} & \texttt{DoS5} & \cellcolor{gray!30}v  & \cellcolor{gray!60}v & v  & \cellcolor{gray!30}v & \cellcolor{gray!60}v & v & \cellcolor{gray!30}v & \cellcolor{gray!60}v & v  \\
\cline{2-11}
 & \texttt{DDoS} & \cellcolor{gray!30}v  & \cellcolor{gray!60}v & v & \cellcolor{gray!30}v  & \cellcolor{gray!60}v & v & \cellcolor{gray!30}v & \cellcolor{gray!60}v & v\\
\cline{2-11}
\hline
\multirow{2}{*}{G(\text{Normal} $\land$ TA $\implies$ X(?))} & \texttt{DoS5} & \cellcolor{gray!30}v  & \cellcolor{gray!60}v &  v  & \cellcolor{gray!30}v & \cellcolor{gray!60}v & v  & \cellcolor{gray!30}v & \cellcolor{gray!60}v & v \\
\cline{2-11}
 & \texttt{DDoS} & \cellcolor{gray!30}v  & \cellcolor{gray!60}v & v & \cellcolor{gray!30}v  & \cellcolor{gray!60}v & v & \cellcolor{gray!30}v & \cellcolor{gray!60}v & v\\
\cline{2-11}
\hline
\end{tabular}
}
\end{table}

\subsubsection{Query-checking.} Table~\ref{table:qc} shows the results of evaluating the state machines obtained from the \texttt{DoS5} and \texttt{DDoS} learning sets against our temporal queries for $R_1$ and $R_2$. Evaluating a temporal query on a state machine either yields a predicate or results in vacuity (v) if the query's antecedent is never met. When the outcome is a predicate, the property obtained by replacing the placeholder in the query with the predicate will hold over the state machine. For example, consider query G\,(\text{Attack} $\land$ TA $\implies$ X\,(?)) in Table~\ref{table:qc}. The results indicate that \system{} never experiences a low-flow attack when in state TA. This is because, for all the state machines, this query is vacuous for low-flow traffic. However, for medium-flow DoS attacks, the query yields W, meaning that when subjected to such attacks in the TA state, the system transitions to the W state.

\subsubsection{Analysis.} The results presented in Tables~\ref{tbl:main-table}(a), \ref{tbl:main-table}(b), and \ref{table:qc} indicate that some temporal properties are  violated and several properties and queries are vacuous.  Below, we discuss the reasons for these violations and vacuities and assess whether the results obtained from different state machines are consistent.

\textbf{Do violations ($\times$) imply defects?} The pass and fail results in Tables~\ref{tbl:main-table}(a) and~\ref{tbl:main-table}(b) show that \system{} reacts to  DoS/DDoS attacks with only medium or high number of flows, ignoring low-flow attacks. Medium-flow attacks prompt \system\ to transition from state S to TW or W, but not to its most critical states, TA or A, which occur only with high-flow attacks.  Dually, \system\  returns to state S only with low-flow normal traffic. In addition, the combination of normal traffic with a high number of flows does not appear in our state machines. This matched the domain expert's intuition, who noted that only attackers (and not normal users) may generate such high numbers of flows. We emphasize that the range for a  ``high'' number of flows is \emph{learned} by our trace abstraction approach and is not a-priori known at the time of trace generation. The fact that the combination of normal traffic with a high number of flows is never generated indicates that our trace abstraction step is able to effectively differentiate between normal users and obvious attackers generating a high number of flows. Furthermore, \system\ remains in the TW and W states for medium-flow normal traffic, indicating that it considers this traffic suspicious.

After discussing these findings, our domain expert confirmed that the system's behaviours are acceptable. In particular, since low-flow DoS and DDoS attacks do not degrade the quality of service for clients, there is no need for the system to react. When recovering from a recent attack, however, it is acceptable for the system to flag medium-sized normal traffic as suspicious. This is because in a real-world deployment, the system does not have an oracle to inform it whether it is still under attack, and it is challenging to distinguish medium-flow normal traffic from DoS and DDoS attacks.

\begin{tcolorbox}[breakable,colback=gray!10!white,colframe=black!75!black]
The property violations identified in Tables~\ref{tbl:main-table}(a) and~\ref{tbl:main-table}(b) do not indicate defects in \system. Rather, they indicate that requirements $R_1$ and $R_2$ need to be refined to explicitly specify the  thresholds for the number of attack and normal flows that the system should treat as suspicious.
\end{tcolorbox}

\textbf{Do vacuities (v) show gaps?} The vacuous cases in Tables~\ref{tbl:main-table}(a) and~\ref{tbl:main-table}(b) are  due to two factors: (1)~Not all numbers of flows are observable in every system state. This is expected since DoS and DDoS attacks typically  start at a low number of flows and gradually escalate. It is typical for a high-flow DoS or DDoS attack to be preceded by a medium-flow phase, which itself follows a low-flow stage. Hence, since a medium-flow attack shifts the state from S to W or TW, a high-flow attack is not observed at state S. (2)~As discussed in Section~\ref{sec:expdesign}, under a DDoS attack, \system{} does not transition to TW and TA states, leading to vacuous outcomes for properties involving these states. Our domain expert confirmed that DDoS attacks involve larger numbers of flows than DoS attacks, causing \system{} to bypass the TW and TA states.

\begin{tcolorbox}[breakable,colback=gray!10!white,colframe=black!75!black]
The observed vacuities are not due to gaps (incompleteness) in the learning sets obtained from our testbed. Rather, vacuity occurs because the physical characteristics of network flows impose certain constraints, such as attacks not starting immediately with a high number of flows but needing to grow over time from a lower number. This indicates that, along with the requirements, the environmental assumptions of \system\ should also be made explicit to help with refining the temporal properties of Tables~\ref{tbl:main-table}(a) and~\ref{tbl:main-table}(b) to avoid vacuity.
\end{tcolorbox}

\textbf{Are the results obtained from different state machines consistent?} Our state machines -- learned based on different configurations, i.e., \texttt{1st}, \texttt{2nd}, and \texttt{Top-2} -- yield highly consistent model-checking and query-checking results. In particular, only $8$\% (3 out of 36 cases) of the model-checking results in Tables~\ref{tbl:main-table}(a) and~\ref{tbl:main-table}(b) are inconsistent. Inconsistency means that, for a given property, learning set, and range for the number of flows, we obtain different results for the state machines built using the \texttt{1st}, \texttt{2nd}, and \texttt{Top-2} configurations.  In all inconsistent cases, out of the three alternative state machines (\texttt{1st}, \texttt{2nd}, and \texttt{Top-2}), two are in agreement, indicating that the inconsistencies can be resolved through a majority voting between the state machines. Similarly, $12.5$\% (3 out of 24 cases) of the query-checking results in Tables~\ref{table:qc} are inconsistent. Nevertheless, in all of these inconsistent cases, the inconsistency can be resolved by weakening the learned predicates through taking their disjunction. The reason this strategy works is that the predicates replace the consequents of their respective temporal queries, and a disjunction (weakened predicate) still ensures the satisfaction of the query. For example,  the query G\,(\text{Attack} $\land$ TW $\implies$ X\,(?)) with medium-flow attack, respectively derives the predicates \hbox{TW $\vee$ TA}, TW, and TW  for configurations \texttt{1st}, \texttt{2nd}, and \texttt{Top-2}. Here, the disjunction of the predicates would be: TW $\vee$ TA.

\begin{tcolorbox}[breakable,colback=gray!10!white,colframe=black!75!black]
State machines obtained by different configurations of \app\ yield highly consistent model-checking and query-checking results, showing the robustness of our trace abstraction approach. Furthermore, inconsistencies can be resolved by using a majority vote for model checking and by taking the disjunction of predicates for query checking.
\end{tcolorbox}

\subsection{Validity Considerations}

\mbox{}\indent\textbf{Internal validity.} To improve internal validity, we implemented measures to minimize the impact of extraneous factors. Specifically, (1) we controlled the trace generation process to prevent external traffic not initiated by our simulations from reaching the local users or the router; (2) we monitored the network during the experiments to ensure the absence of anomalies that might have arisen due to events beyond our control; and (3) to construct each dataset, we repeated the trace generation process five times and combined the results, \hbox{thereby mitigating the effects of random variability.}

\textbf{External validity.} 
While we believe our approach should generalize to other types of systems with numeric time-series inputs and outputs, we note that our evaluation was conducted on a single system in the domain of network intrusion detection. To more conclusively examine the generalizability of our approach and improve external validity, further experimentation with other systems, such as cyber-physical systems, would be necessary.

%% file: Sections/relwork.tex
\section{Related Work}
\label{sec:relwork}
In this section, we compare our work with relevant strands in three
areas described below: 

\textit{\bfseries Automata learning and verification for network protocols.} Muskardin et al.~\cite{muskardin2022active} employ AALpy -- the same tool we use in this paper for automata learning -- to compare passive and active learning for network protocols~\cite{PferscherA21}. They observe that active learning: (1) is time-consuming, especially when the system under learning  needs resetting for each new input/output trace; (2) requires a fault-tolerant learning setup for interaction, which can be complex and limit practicality; and (3) involves significant interaction with the system under learning, incurring high costs due to the potential for losses or delays. Based on these observations, Muskardin et al. argue that passive learning is a more efficient alternative for real-world problems if one can have a diverse and yet sparse dataset of system behaviours. Our work follows the same rationale for choosing passive learning over active learning. To improve learning-set diversity and coverage, we used randomization by rerunning the data generation loop of Algorithm~\ref{alg:approach} multiple times, further ensuring that we captured all known system states. The resulting learning sets are relatively small (with an average trace length of \~1500), thus remaining amenable to effective passive learning.

Fiterau et al.~\cite{fiteruau2016combining} employ automata learning to derive behavioural models for TCP protocol components and apply model checking to verify their conformity with Request for Comments (RFCs). Our approach to learning differs from theirs in two main respects: (1) they do not address numeric systems, making their abstractions unsuitable for our context, and (2) they use active learning, whereas we use passive learning. Both our work and Fiterau et al.'s employ model checking to verify the resulting models. However, the nature of the properties of interest differs: while they examine interactions in TCP network protocol components, we assess the behaviours of an entire system (a network router) in its deployment environment.

\textit{\bfseries Model mining for intrusion detection systems.} ML technologies have become crucial for enhancing automated intrusion detection systems. However, as observed by Shahraki et al.~\cite{ShahrakiTH21}, ML techniques used for network-traffic monitoring and analysis (including for our use case in this paper) have thus far been heavyweight and primarily tailored to enterprise networks. Little attention has been given to lighter-weight techniques that would be suitable for the needs of small networks or resource-parsimonious networking platforms, such as those prevalent in the market where our industry partner operates. Another issue related to the application of ML for intrusion detection is the lack of interpretability, as humans often struggle to comprehend the deep operational layers of ML-based intrusion detection systems~\cite{WangZYW20}.

We are not the first to explore the application of passive learning to mine interpretable models for network intrusion detection systems. Cao et al.~\cite{cao2022} use passive learning to construct probabilistic state machines for detecting network anomalies based on features such as protocol, bytes sent, and flow duration. We differ from Cao et al. in two key aspects. First, our work focuses on router firewall behaviours, emphasizing state transitions (Safe, Warning, Alert) based on network flows, whereas Cao et al. target anomaly detection in Kubernetes clusters. Our use case differs from theirs, and the solutions are not interchangeable. Second, whereas Cao et al. use clustering for numeric-range partitioning, we employ decision-tree learners. Noting that our experimental setup allows for control over attack and non-attack scenarios without manual effort or compromising accuracy in labelling, decision trees are advantageous over clustering due to their better interpretability and resilience to outliers and dataset imbalances.

\textit{\bfseries Supervised rule mining for numeric systems.} Our research relies on supervised rule mining to improve the abstraction of behavioural models for numeric systems. While we are not aware of prior research that employs supervised rule mining for a similar application, recent work on rule mining for numeric systems has inspired our approach. Notably, Jodat et al.~\cite{JodatNSS23, TOSEM} propose a method for combining machine learning and adaptive random testing to identify test inputs leading to non-robust and potentially failing system behaviours. Our current research was conducted with the same industry partner as Jodat et al. (namely,  RabbitRun Technologies). However, this prior work focuses on a different aspect of the partner's system, specifically controlling the flow of network traffic (traffic shaping). In short, this earlier research neither concerns learning behavioural models nor addresses intrusion detection.

%% file: Sections/lessons.tex
\section{Lessons Learned and Future Work}
\label{sec:lessons}
Below, we reflect on the lessons learned from the development of \app. 
\emph{(1) For systems with time-series inputs and outputs, automata learning produces effective, interpretable behavioural models. While interpretable statistical learning is effective at deriving static abstractions from data, it is not as effective at capturing  temporal behaviours encoded in time-series data.}  Interpretable ML methods~\cite{Interptlbook}, such as decision trees and decision rules, are effective in identifying predicates that explain the relationship between system inputs and system states. However, they are inadequate in capturing temporal relationships between states and in understanding how system inputs might trigger change of states. Our research shows that: (1) automata learning, due to its ability to capture temporal behaviours effectively, proves useful for mitigating the shortcomings of statistical learning, and (2) to overcome the limitations of automata learning in abstracting data, one can increase the level of abstraction in traces first before attempting to learn automata.

\emph{(2)~Automata learning is useful for analyzing and gaining a better understanding of cyber-intrusion detection systems.} There is a lack of industrial case studies on the automated derivation of behavioural models for cyber-intrusion detection systems. Our work highlights some important contextual factors related to the construction of behavioural models for such systems, notably the numeric and time-series nature of the inputs and outputs of these systems, as well as the lack of amenability to active learning techniques due to the difficulty of building query-and-response loops. An important lesson learned from our work is the feasibility of automata learning for cyber-intrusion detection systems through an explicit treatment of time-series \hbox{numeric data and the use of passive learning.}

\vspace*{.5em}
\textbf{Data Availability.} We publicly share our scripts, trace creation and abstraction routines, along with all experimental data~\cite{neayoughi2024mela}.

\vspace*{.5em}\textbf{Acknowledgements.} We gratefully acknowledge the financial support received from NSERC of Canada through the Discovery, Discovery Accelerator, and Alliance programs.